\documentclass[reprint,aps,prd,onecolumn,notitlepage,showpacs,nofootinbib,preprintnumbers,superscriptaddress]{revtex4-1}
\usepackage{amsmath}
\usepackage{amsfonts}
\usepackage{amssymb}
\usepackage{latexsym}
\usepackage{enumerate}
\usepackage{color,xcolor}
\usepackage{graphicx}
\usepackage{bm}
\usepackage{epsfig}
\usepackage[english]{babel}
\usepackage{hyperref}
\usepackage{times}
\usepackage{comment}
\usepackage{epstopdf}

\def\N{\mathcal{N}}
\def\B{\mathcal{B}}
\def\dom{d\Omega^2}
\def\rmin{r_\mathrm{min}}
\def\rhom{\rho_\mathrm{min}}
\def\rph{r_\mathrm{ph}}
\def\rhoph{\rho_\mathrm{ph}}
\def\ro{r_\mathrm{o}}
\def\rhoo{\rho_\mathrm{o}}
\def\Rsh{R_\mathrm{sh}}
\def\bph{b_\mathrm{ph}}
\def\ash{\alpha_\mathrm{sh}}
\def\cs{c_\mathrm{S}}
\def\vmin{V_\mathrm{min}}
\def\vph{V_\mathrm{ph}}

\begin{document}

\title{Shadow of the wormholelike static aether solution}
\author{Yanni Zhu}
\author{Towe Wang}
\email[Electronic address: ]{twang@phy.ecnu.edu.cn}
\affiliation{Department of Physics, East China Normal University,\\
Shanghai 200241, China\\ \vspace{0.2cm}}
\date{\today\\ \vspace{1cm}}
\begin{abstract}
Recently, an analytical solution in Einstein-aether theory was presented explicitly in isotropic coordinates. It is characterized by a mass parameter $m$ and a combined coupling constant $c_{14}$. By a coordinate transformation, we verify this solution is equivalent to the previously known static aether solution. Assuming photons couple minimally to aether and gravity, we investigate the photon sphere and shadow of the solution by varying $c_{14}$. Results are compared with those of the Schwarzschild black hole.
\end{abstract}


\maketitle




\section{Introduction}\label{sect-intro}
The shadow cast is a robust feature of black holes \cite{Synge:1966okc,Bardeen:1973,Luminet:1979nyg} surrounded by optically thin and geometrically thick emission regions \cite{Falcke:1999pj,Bronzwaer:2021lzo}. It was imaged successfully for the first time by the Event Horizon Telescope \cite{EventHorizonTelescope:2019dse,EventHorizonTelescope:2019pgp,EventHorizonTelescope:2019ggy} in the center of galaxy M87. This opened a new window to probe the geometry of spacetime near black holes. In Einstein theory with a 4-dimensional spacetime, the geometry of vacuum black holes, described by solutions with regular event horizons, is subject to uniqueness theorems \cite{Israel:1967wq,Robinson:1975bv}. Therefore, one can test Einstein's theory by confronting predictions of the Schwarzschild or Kerr solution against the observational data of shadow \cite{Bambi:2019tjh,Vagnozzi:2019apd,Psaltis:2018xkc}.

Einstein-aether theory is a vector-tensor theory that breaks Lorentz invariance by coupling a unit timelike vector field (the aether field) to the metric at every point in spacetime \cite{Jacobson:2000xp,Jacobson:2007veq}. In this theory, the shadow of a numerical black hole \cite{Eling:2006ec} has been investigated in Ref. \cite{Zhu:2019ura}. Among analytical solutions, shadows have been studied in Ref. \cite{Khodadi:2020gns} for spherical black holes \cite{Berglund:2012bu}, in Refs. \cite{Zhu:2019ura,Liu:2021yev} for spherical charged black holes \cite{Ding:2015kba}, and in Refs. \cite{Khodadi:2020gns,Zhu:2019ura,Liu:2021yev} for a slowly rotating black hole \cite{Barausse:2015frm}.

Interestingly, in all of these black hole solutions \cite{Eling:2006ec,Berglund:2012bu,Ding:2015kba,Barausse:2015frm}, the aether field is not aligned with the horizon-generating Killing vector. The reason seems to be obvious: the aether field $u^a$ is timelike everywhere but the Killing field $\xi^a$ becomes lightlike on its horizon \cite{Eling:2006ec}. However, there is a loophole in this reasoning. When the aether field is proportional to the Killing field, $u^a=f(r)\xi^a$, it is still possible to satisfy $g_{ab}u^au^b=-1$ in the full spacetime if $g_{ab}\xi^a\xi^b=-1/f^2\rightarrow0$ on the Killing horizon and nonzero elsewhere. One close example is a wormholelike solution derived by Eling and Jacobson (EJ) in Ref. \cite{Eling:2006df}, referred to also as a static aether solution therein. In this solution, the unit timelike aether field is aligned with the timelike Killing field, and the latter field becomes lightlike on a hypersurface. Unfortunately, this hypersurface is not a regular horizon, because a curvature component gets singular on it \cite{Eling:2006df}. Between the curvature singularity and the spatial infinity, the solution has a throat with a minimal area sphere as happens in wormholes. All of the analyses in Ref. \cite{Eling:2006df} were done in Schwarzschild coordinates, but the metric cannot be given explicitly. Prior to Ref. \cite{Eling:2006df}, the asymptotic weak field limit of spherical static solutions has been studied perturbatively in Ref. \cite{Eling:2003rd} in isotropic coordinates.

Recently, in Einstein-aether theory, Oost, Mukohyama and Wang (OMW) \cite{Oost:2021tqi} derived an exact static solution in a closed form with the aether field parallel to the timelike Killing field. The solution is asymptotically flat, written in isotropic coordinates and characterized by two parameters: the mass $m$ as an integration constant, and a combined coupling $c_{14}$ preset by the action. Setting $c_{14}=2$, the metric reduces to the Schwarzschild solution in Einstein theory in isotropic coordinates, but the aether field does not vanish. When $c_{14}\neq2$, the solution describes a marginally trapped finite-size throat smoothly connecting two untrapped patches: one patch has a singularity at a finite proper distance, and the other patch is asymptotically flat at the infinite proper distance. The aether configuration and the spacetime geometry are reminiscent of the wormholelike static aether solution derived in Ref. \cite{Eling:2006df}. As mentioned in Ref. \cite{Oost:2021tqi}, they must be the same solution in different coordinate systems. In the present paper, we will put this statement on firmer ground by transforming the solution from isotropic coordinates to Schwarzschild coordinates.

Observationally, the shadow cast is an important window to black hole astrophysics focusing on physical black hole solutions in a well-established gravity theory in astrophysical environments. Theoretically, the shadow provides us with a picture illustrating the motion of light rays in many gravitational solutions to most gravity theories. In both situations, the computation is facilitated a lot by an explicit analytical form of line element. In the spirit of the cosmic censorship conjecture \cite{Penrose:1969pc}, the wormholelike static aether solution is physically undesirable, because it has a curvature singularity inside the throat but the throat is not a Killing horizon. This is a naked null singularity. Investigations in Refs. \cite{Bambi:2008jg,Ortiz:2015rma,Shaikh:2018lcc,Joshi:2020tlq} showed that some naked singularities can also produce shadows similar to but slightly different from black holes. Therefore, it becomes promising and interesting to test the cosmic censorship conjecture with the observed shadow cast, which is one motivation of this paper. Neglecting light rays emanated from the marginally trapped hypersurface, the photon sphere and the shadow cast can be computed for the static aether solution in a similar way to the Schwarzschild black hole. The influence of parameters on the motion of light rays can be depicted by the shadow image more obviously than the trajectory graph.

The rest of the paper is organized as follows: In Sec. \ref{sect-rev}, we will give a brief review for the EJ solution \cite{Eling:2006df} and the OMW solution \cite{Oost:2021tqi} in Einstein-aether theory, and set up our convention of notations. In Sec. \ref{sect-equ}, we will transform the OMW solution from isotropic coordinates to Schwarzschild coordinates in Sec. \ref{subsect-OMW} and reproduce the resulting solution with the EJ solution in Sec. \ref{subsect-EJ}. Some special limits will be discussed in Sec. \ref{subsect-lim}. In Sec. \ref{sect-shad}, we will compute the radii of the photon sphere and the shadow cast of the wormholelike static aether solution in Sec. \ref{subsect-rad}, simulate light trajectories and intensity images in Sec. \ref{subsect-int} and evaluate the volume inside the throat in Sec. \ref{subsect-vol}. Sec. \ref{sect-con} will be devoted to our conclusion.

\section{Brief review and convention of notations}\label{sect-rev}
The action for Einstein-aether theory \cite{Jacobson:2000xp,Jacobson:2007veq} is
\begin{equation}\label{act}
S=\frac{1}{16\pi G}\int dx^4\sqrt{-g}\left[R-K^{ab}_{~~~mn}\nabla_au^m\nabla_bu^n+\lambda\left(g_{ab}u^au^b+1\right)\right],
\end{equation}
in which $\nabla$ denotes the covariant derivative, $a,b=0,1,2,3$, and the tensor
\begin{equation}
K^{ab}_{~~~mn}=c_1g^{ab}g_{mn}+c_2\delta^a_m\delta^b_n+c_3\delta^a_n\delta^b_m-c_4u^au^bg_{mn}.
\end{equation}
Here $c_i$ ($i=1,2,3,4$) are dimensionless coupling constants. In Eq. \eqref{act}, $\lambda$ is the Lagrangian multiplier to ensure that the aether field $u^a$ is a unit timelike vector.

Throughout this paper, we assume photons and other matter couple minimally to aether and gravity, so the speed of matter is limited by the light speed $c$. From now on, we will adopt the geometrized unit system with $G=c=1$ and use the notations $c_{ij}=c_i+c_j$, $c_{ijk}=c_i+c_j+c_k$. According to Refs. \cite{Oost:2021tqi,Oost:2018tcv,Zhang:2020too}, current theoretical and observational constraints impose $0\leq c_{14}\leq2.5\times10^{-5}$, thus we will always take $c_{14}$ as a small nonnegative parameter.

The static aether or wormholelike solution written in Schwarzschild coordinates ($t,\rho,\theta,\phi$) in Ref. \cite{Eling:2006df} takes the form
\begin{eqnarray}\label{EJ}
\nonumber ds^2&=&-\N(\rho)dt^2+\B(\rho)d\rho^2+\rho^2\dom,\\
u^a\partial_a&=&\N(\rho)^{-1/2}\partial_t
\end{eqnarray}
with the standard metric on the unit sphere $\dom=d\theta^2+\sin^2\theta d\phi^2$. To avoid conflict of notation, we use $\N$, $\B$ and $\rho$ here instead of $N$, $B$ and $r$ in Ref. \cite{Eling:2006df}. The Schwarzschild radial coordinate $\rho$ will be called the areal radial coordinate or the areal radius, because the area of a sphere centered on the origin is $4\pi \rho^2$. In Eq. \eqref{EJ}, with the help of an implicit function $Y(\rho)$ of $\rho$ decided by
\begin{equation}\label{rhoY}
\frac{\rhom}{\rho}=\left(\frac{Y}{Y-Y_{-}}\right)\left(\frac{Y-Y_{-}}{Y-Y_{+}}\right)^{\frac{1}{2+Y_{+}}},
\end{equation}
the dependence of $\N(\rho)$ and $\B(\rho)$ on $\rho$ was given in Ref. \cite{Eling:2006df} as
\begin{equation}\label{NBY}
\N(\rho)=\left(\frac{1-Y/Y_{-}}{1-Y/Y_{+}}\right)^{\frac{-Y_{+}}{2+Y_{+}}},~~~~\B(\rho)=1+Y+\nu Y^2.
\end{equation}
In the above, the minimum radius $\rhom$ is an integration constant, $Y_{\pm}=(-1\pm\sqrt{1-4\nu})/(2\nu)$ are two roots of $\B$, and the symbol $\nu=c_{14}/8\geq0$. As pointed out by Ref. \cite{Eling:2006df}, when $\nu>0$, the metric \eqref{EJ} has a curvature singularity at $Y=Y_{-}$ and a throat (a minimal area sphere) at $\rho=\rhom$.

In Ref. \cite{Oost:2021tqi}, a new form of exact solution to Einstein-aether theory was presented analytically in isotropic coordinates ($t,r,\theta,\phi$),
\begin{eqnarray}\label{OMW}
\nonumber ds^2&=&-\left(\frac{1-\frac{m}{2r}}{1+\frac{m}{2r}}\right)^qdt^2+\frac{\left(1+\frac{m}{2r}\right)^{q+2}}{\left(1-\frac{m}{2r}\right)^{q-2}}\left(dr^2+r^2\dom\right),\\
u^a\partial_a&=&\left(\frac{1+\frac{m}{2r}}{1-\frac{m}{2r}}\right)^{q/2}\partial_t,
\end{eqnarray}
in which $m$ is an integration constant and
\begin{equation}\label{q}
q=2\left(\frac{2}{2-c_{14}}\right)^{1/2}\geq2.
\end{equation}
As demonstrated in Ref. \cite{Oost:2021tqi}, when $q>2$, the metric \eqref{OMW} has a curvature singularity and a marginally trapped throat. In isotropic coordinates, the curvature singularity occurs at $r=m/2$, which is a null singularity \cite{Eling:2006df}, and the throat sits at
\begin{equation}\label{rmin}
r=\rmin=\frac{m}{4}\left(q+\sqrt{q^2-4}\right).
\end{equation}

For brevity in this paper, the patch $m/2<r\leq\rmin$, i.e. the spacetime between the throat and the curvature singularity will be referred to as the inner patch, while the patch $r\geq\rmin$, i.e. the spacetime from the throat to asymptotic infinity will be referred to as the outer patch.

\section{Equivalence of EJ and OMW solutions}\label{sect-equ}
Except for that they are written in different coordinate systems, both EJ and OMW solutions \cite{Eling:2006df,Oost:2021tqi} are derived under the following conditions:
\begin{enumerate}[(i)]
\item The metric is static, spherically symmetric and asymptotically flat.
\item The aether is aligned with the timelike Killing vector.
\end{enumerate}
Therefore, in principle they should be the same solution in different coordinates. That is to say, they should be related to each other by a transformation of coordinates.

In this section, we will verify the equivalence of the two solutions in two steps: firstly transforming the OMW solution from isotropic coordinates to Schwarzschild coordinates via the coordinate transformation
\begin{equation}\label{rhor}
\rho=\frac{\left(1+\frac{m}{2r}\right)^{q/2+1}}{\left(1-\frac{m}{2r}\right)^{q/2-1}}r
\end{equation}
in Sec. \ref{subsect-OMW}, and secondly reproducing the resulting solution with the EJ solution by replacement
\begin{equation}\label{Yr}
Y=\frac{4qmr}{4r^2-2qmr+m^2}
\end{equation}
in Sec. \ref{subsect-EJ}. As a byproduct, we obtain an alternative form of static aether solution in Schwarzschild coordinates, which involves $r$ instead of $Y$ as an implicit function of $\rho$. Some of its special limits will be discussed in Sec. \ref{subsect-lim}.

\subsection{Transforming OMW solution}\label{subsect-OMW}
Starting with the analytic solution \eqref{OMW}, we can transform it into Schwarzschild coordinates ($t,\rho,\theta,\phi$) via the coordinate transformation \eqref{rhor} so that it takes the expected form \eqref{EJ}. Comparing Eqs. \eqref{EJ} and \eqref{OMW}, it is easy to read off
\begin{equation}\label{NBr}
\N(\rho)=\left(\frac{1-\frac{m}{2r}}{1+\frac{m}{2r}}\right)^q,~~~~\B(\rho)d\rho^2=\frac{\left(1+\frac{m}{2r}\right)^{q+2}}{\left(1-\frac{m}{2r}\right)^{q-2}}dr^2.
\end{equation}
Differentiation of Eq. \eqref{rhor} yields
\begin{equation}
\frac{1}{\rho}d\rho=\frac{4r^2-2qmr+m^2}{r(4r^2-m^2)}dr,
\end{equation}
inserting which into Eq. \eqref{NBr} we obtain
\begin{equation}\label{Br}
\B(\rho)=\left(\frac{4r^2-m^2}{4r^2-2qmr+m^2}\right)^2.
\end{equation}
In this way, $\N(\rho)$ and $\B(\rho)$ are expressed as composite functions of $r$ of $\rho$, where the inner function $r$ is an implicit function of $\rho$ dictated by Eq. \eqref{rhor}.

There is a pitfall here. As indicated by Figure 1 in Ref. \cite{Oost:2021tqi}, the areal radius $\rho$ is not a monotonic function of $r\in[m/2,\infty)$ unless $q=2$. In consequence, the reverse function of Eq. \eqref{rhor} does not exist in the full region $r>m/2$, so functions $\N(\rho)$ and $\B(\rho)$ are ill defined. However, we note that Eq. \eqref{rhor} decreases monotonically in the interval $(m/2,\rmin]$ and increases monotonically in the region $[\rmin,\infty)$. To remedy the pitfall, when covering the spacetime \eqref{OMW} by Schwarzschild coordinates, we divide the spacetime into two patches corresponding to $m/2<r\leq\rmin$ and $r\geq\rmin$ respectively. Then making use of Eqs. \eqref{rhor}, \eqref{NBr} and \eqref{Br}, functions $\N(\rho)$ and $\B(\rho)$ can be defined well in each patch separately.

\subsection{Some special limits}\label{subsect-lim}
The two patches are smoothly connected at the throat $r=\rmin$ \cite{Oost:2021tqi}, where the areal radius \eqref{rhor} takes the minimum value
\begin{equation}\label{rhom}
\rhom=\frac{m}{2}\sqrt{q^2-4}\left(\frac{q+2}{q-2}\right)^{q/4}
\end{equation}
and Eq. \eqref{EJ} tends to
\begin{eqnarray}\label{Schm}
\nonumber ds^2&\rightarrow&-\left(\frac{q-2}{q+2}\right)^{q/2}dt^2+\frac{\rhom}{2(\rho-\rhom)}d\rho^2+\rhom^2\dom,\\
u^a\partial_a&\rightarrow&\left(\frac{q+2}{q-2}\right)^{q/4}\partial_t.
\end{eqnarray}
In Sec. \ref{subsect-EJ}, we will prove that the parameter $\rhom$ in Eq. \eqref{rhoY} is equal to the one in Eq. \eqref{rhom}.

At the singularity $r\rightarrow m/2$, the areal radial coordinate $\rho\rightarrow\infty$, and Eq. \eqref{EJ} approaches
\begin{eqnarray}\label{Schl}
\nonumber ds^2&\rightarrow&-\left(\frac{2m}{\rho}\right)^{2q/(q-2)}dt^2+\left(\frac{4}{2-q}\right)^2\left(\frac{2m}{\rho}\right)^{4/(q-2)}d\rho^2+\rho^2\dom,\\
u^a\partial_a&\rightarrow&\left(\frac{\rho}{2m}\right)^{q/(q-2)}\partial_t.
\end{eqnarray}
At the spatial infinity $r\rightarrow\infty$, the areal radius $\rho\rightarrow\infty$, and the metric Eq. \eqref{EJ} is asymptotically flat,
\begin{eqnarray}\label{Schr}
\nonumber ds^2&\rightarrow&-dt^2+d\rho^2+\rho^2\dom,\\
u^a\partial_a&\rightarrow&\partial_t.
\end{eqnarray}

In the special case $c_{14}=0$, i.e. $q=2$, line element \eqref{OMW} coincides with the Schwarzschild solution in isotropic coordinates in Einstein theory, but the aether field does not vanish. In Schwarzschild coordinates, we can now write down the solution explicitly,
\begin{eqnarray}\label{Sch2}
\nonumber ds^2&=&-\left(1-\frac{2m}{\rho}\right)dt^2+\left(1-\frac{2m}{\rho}\right)^{-1}d\rho^2+\rho^2\dom,\\
u^a\partial_a&=&\left(1-\frac{2m}{\rho}\right)^{-1/2}\partial_t.
\end{eqnarray}
There is an obstacle to analytically continuing this solution to the interior of the Schwarzschild black hole, because the aether field in Eq. \eqref{Sch2} is ill defined for $\rho<2m$. We leave this as an open problem for future research and in the present paper restrict our attention to the exterior of Schwarzschild black hole.

\subsection{Rewriting EJ solution}\label{subsect-EJ}
In Schwarzschild coordinates, the metric and aether field are determined by functions $\N(\rho)$ and $\B(\rho)$ completely. The two functions are given by Eqs. \eqref{rhoY}, \eqref{NBY} in EJ solution, and by Eqs. \eqref{rhor}, \eqref{NBr}, \eqref{Br} in Sec. \ref{subsect-OMW}. In this subsection, we will replace $Y$ with Eq. \eqref{Yr} and demonstrate that Eqs. \eqref{rhoY}, \eqref{NBY} are equivalent to Eqs. \eqref{rhor}, \eqref{NBr}, \eqref{Br}.

For preparation, we reverse Eq. \eqref{q} to get
\begin{equation}\label{nuq}
\nu=\frac{1}{4}-\frac{1}{q^2},
\end{equation}
then the value of $Y_{\pm}$ given below Eq. \eqref{NBY} can be reexpressed as
\begin{equation}\label{Ypm}
Y_{\pm}=\frac{-2q}{q\pm2}.
\end{equation}
Substituting Eqs. \eqref{Yr}, \eqref{nuq}, \eqref{Ypm} into Eqs. \eqref{rhoY}, \eqref{NBY}, we find after some straightforward algebras,
\begin{eqnarray}
\frac{\rhom}{\rho}&=&\frac{m\sqrt{q^2-4}}{2r}\left(\frac{q+2}{q-2}\right)^{q/4}\frac{\left(1-\frac{m}{2r}\right)^{q/2-1}}{\left(1+\frac{m}{2r}\right)^{q/2+1}},\\
\N(\rho)&=&\left(\frac{1-\frac{m}{2r}}{1+\frac{m}{2r}}\right)^q,\\
\B(\rho)&=&\left(\frac{4r^2-m^2}{4r^2-2qmr+m^2}\right)^2.
\end{eqnarray}
They are apparently the same as Eqs. \eqref{rhor}, \eqref{NBr}, \eqref{Br} if we identify $\rhom$ with Eq. \eqref{rhom}.

As an aside, we note that Eq. \eqref{Yr} is equivalent to
\begin{equation}\label{rY}
r=\frac{m}{2}\left\{\frac{1}{\sqrt{1-4\nu}}\left[\frac{Y(\rho)+2}{Y(\rho)}\right]+\frac{2}{\sqrt{1-4\nu}}\left[\frac{\nu Y(\rho)^2+Y(\rho)+1}{Y(\rho)^2}\right]^{1/2}\right\}
\end{equation}
through which one can write a similar but more tedious proof.

Recalling that Eqs. \eqref{rhor}, \eqref{NBr}, \eqref{Br} are derived from the OMW solution by a coordinate transformation, so we can conclude the OMW solution is equivalent to the EJ solution as promised.

\section{Shadow cast}\label{sect-shad}
The line element Eq. \eqref{OMW}, which describes the wormholelike static aether spacetime, takes an explicit analytical form. It provides us with a good starting point to study the light trajectories, the photon sphere and the shadow cast of this spacetime. In the current section, we will do such a study by ignoring all influences from the patch $m/2<r\leq\rmin$, i.e. the spacetime between the throat and the curvature singularity. The radii of the photon sphere and the shadow cast will be computed in Sec. \ref{subsect-rad}, while light trajectories and intensity images will be simulated in Sec. \ref{subsect-int}. To justify the ignoring of influences from the inner patch, we will evaluate the volume inside the throat in Sec. \ref{subsect-vol}.

For brevity and generality, let us write the line element of a static spherical spacetime in the general form
\begin{equation}\label{metric-sph}
ds^2=-A(r)dt^{2}+B(r)dr^{2}+D(r)\dom.
\end{equation}
In every subsection, general formulas will be written in terms of $A$, $B$ and $D$ and then applied to the specific case Eq. \eqref{OMW}, which corresponds to
\begin{equation}\label{ABD}
A(r)=\left(\frac{1-\frac{m}{2r}}{1+\frac{m}{2r}}\right)^q,~~~~B(r)=\frac{D(r)}{r^2}=\frac{\left(1+\frac{m}{2r}\right)^{q+2}}{\left(1-\frac{m}{2r}\right)^{q-2}}.
\end{equation}
To make full use of the explicit expression \eqref{OMW}, calculations will be done in isotropic coordinates. But for comparison with the familiar Schwarzschild black hole, final results and figures are presented in terms of Schwarzschild coordinates. We will assume a distant observer with a radial coordinate $\ro\rightarrow\infty$. In the special case $q=2$, all results in this section reduce to those of the Schwarzschild black hole.

\subsection{Radii}\label{subsect-rad}
The shadow cast of a spherical black hole or its mimickers has the shape of a circular disk, whose edge is a lensed image of the outermost photon sphere. The radius of the sphere $\rph$ is the largest root of the equation \cite{Perlick:2015vta,Perlick:2021aok}
\begin{equation}\label{rph}
\frac{d}{dr}\left[\frac{D(r)}{A(r)}\right]=0.
\end{equation}
Observed at a radius of $\ro$, the opening angle of the shadow $\ash$ satisfies \cite{Perlick:2015vta,Perlick:2021aok}
\begin{equation}\label{ash}
\sin^2\ash=\frac{D(\rph)A(\ro)}{A(\rph)D(\ro)},
\end{equation}
and the radius squared of the shadow is
\begin{equation}\label{bsh}
\Rsh^2=\rhoo^2\sin^2\ash=\frac{D(\rph)A(\ro)}{A(\rph)}.
\end{equation}

\begin{figure}
\centering
\includegraphics[width=0.45\textwidth]{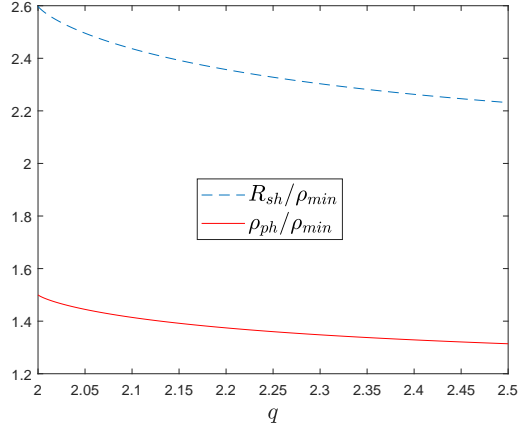}\\
\caption{Radii of the photon sphere and the shadow normalized by $\rhom$. The radius of photon sphere is represented by a red solid curve following Eq. \eqref{rphae}, while the radius of shadow is denoted by a blue dashed curve according to Eq. \eqref{bshae}. As $q$ increases, both curves go down.}\label{fig-rad}
\end{figure}

When $A(r)$ and $D(r)$ take the form \eqref{ABD}, the largest root of Eq. \eqref{rph} is
\begin{equation}\label{rphae}
r=\rph=\frac{m}{2}\left(q+\sqrt{q^2-1}\right).
\end{equation}
Substituting it into Eq. \eqref{rhor}, we find the areal radius of photon sphere
\begin{eqnarray}\label{rhophae}
\nonumber\rhoph&=&m\sqrt{q^2-1}\left(\frac{q+1}{q-1}\right)^{q/4}\\
&=&2\rhom\sqrt{\frac{q^2-1}{q^2-4}}\left[\frac{(q+1)(q-2)}{(q-1)(q+2)}\right]^{q/4}.
\end{eqnarray}
To obtain the areal radius of shadow observed at spatial infinity $\ro\rightarrow\infty$, we make use of Eqs. \eqref{ABD} and \eqref{rphae} to evaluate Eq. \eqref{bsh}. The result can be simplified to
\begin{eqnarray}\label{bshae}
\nonumber\Rsh&=&m\sqrt{q^2-1}\left(\frac{q+1}{q-1}\right)^{q/2}\\
&=&2\rhom\sqrt{\frac{q^2-1}{q^2-4}}\left[\frac{(q+1)^2(q-2)}{(q-1)^2(q+2)}\right]^{q/4}.
\end{eqnarray}
Fixing the areal radius of throat $\rhom$ and varying the parameter $q$, we plotted our results in Fig. \ref{fig-rad}. From the figure we can see both radii decrease monotonically as $q$ increases in the interval $2\leq q\leq2.5$.

\subsection{Trajectories and intensity}\label{subsect-int}
In this subsection, we are interested in the image of the shadow cast formed by light rays arriving at the observer. In virtue of the spherical symmetry, we can restrict our study to clockwise light rays on the equatorial plane, whose orbit equation is \cite{Perlick:2015vta,Perlick:2021aok}
\begin{equation}\label{orbitr}
\frac{dr}{d\phi}=-\sqrt{\frac{D(r)}{B(r)}\left[\frac{D(r)}{b^2A(r)}-1\right]}.
\end{equation}
Here $b$ is the impact parameter, its critical value \cite{Perlick:2021aok} can be expressed as
\begin{equation}\label{bc}
\bph=\sqrt{\frac{D(\rph)}{A(\rph)}}=m\frac{\left(q+1\right)^{(q+1)/2}}{\left(q-1\right)^{(q-1)/2}}.
\end{equation}
In the image seen at an areal radius $\rhoo$, photons with the same impact parameter $b$ form a circle of radius squared \cite{Li:2021btf}
\begin{equation}\label{R}
R^2=\rhoo^2\sin^2\alpha=b^2A(\ro).
\end{equation}
The intensity image of the shadow cast is sensitive to the motion of radiating gas which plays the role of light source \cite{Narayan:2019imo,Zeng:2020vsj}. Here we consider the simple case of gas at rest with a specific emissivity
\begin{equation}\label{jA}
j(\nu,r)\propto\frac{\delta(\nu-\nu_0)}{D(r)},
\end{equation}
then the intensity at the point $(x,y)$ of the observed image is
\begin{equation}\label{I}
I(x,y)=\int_{\mathrm{ray}}\frac{A(r)^{3/2}}{D(r)}\sqrt{B(r)+D(r)\left(\frac{d\phi}{dr}\right)^2}dr
\end{equation}
up to an overall normalization factor.

\begin{figure}
\centering
\includegraphics[width=0.45\textwidth]{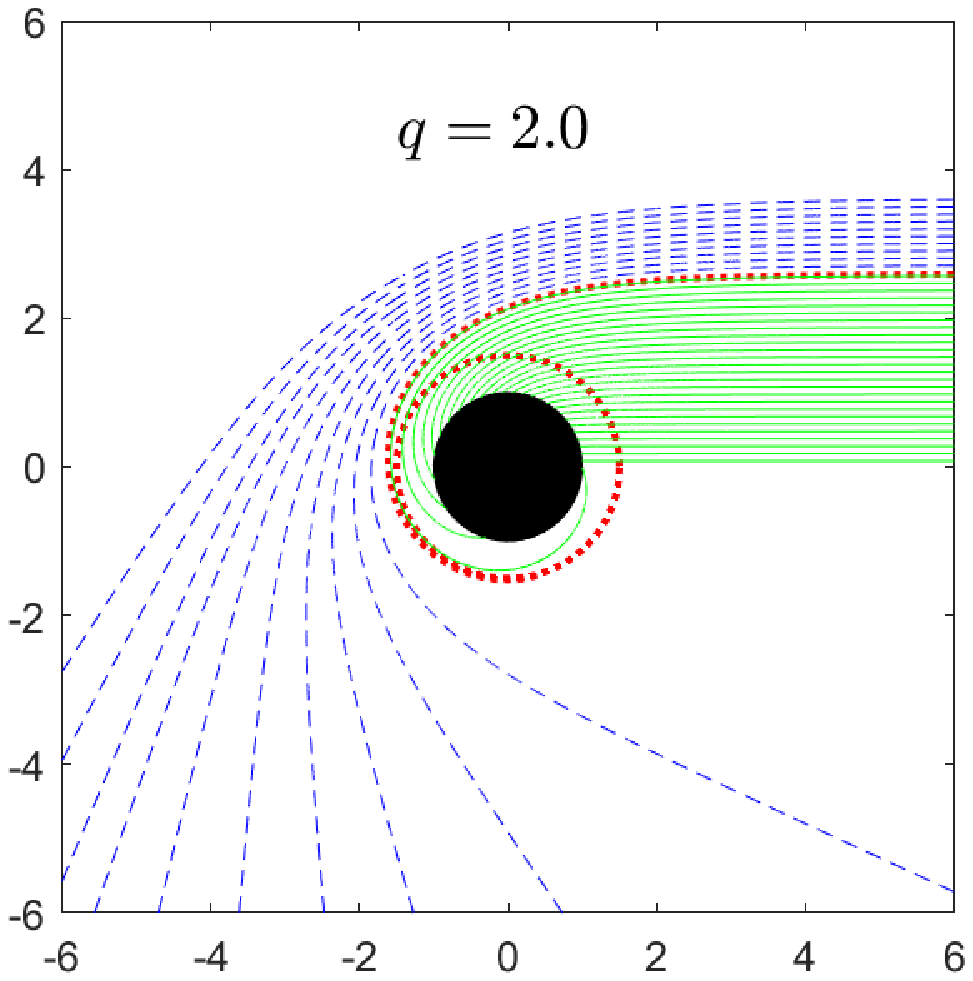}\includegraphics[width=0.45\textwidth]{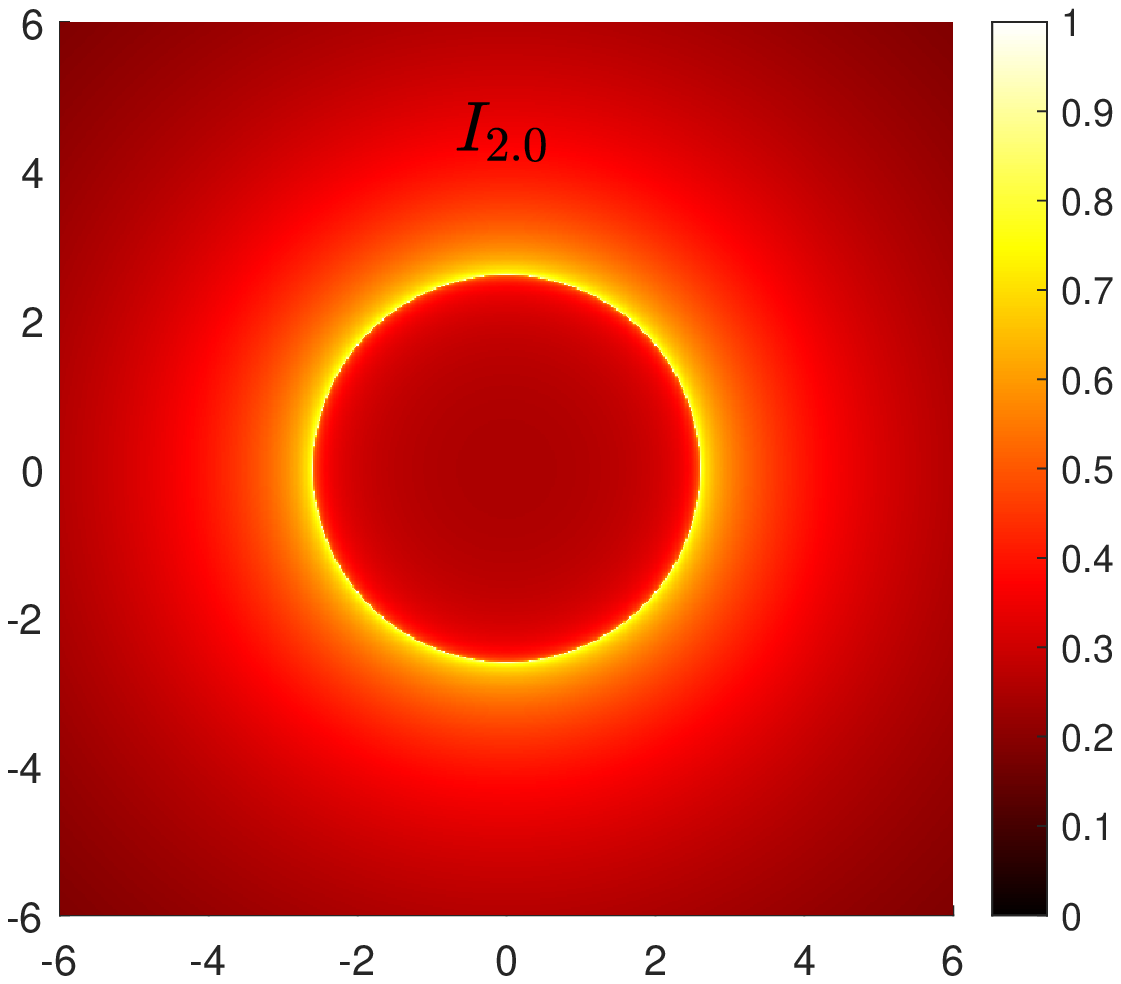}\\
\includegraphics[width=0.45\textwidth]{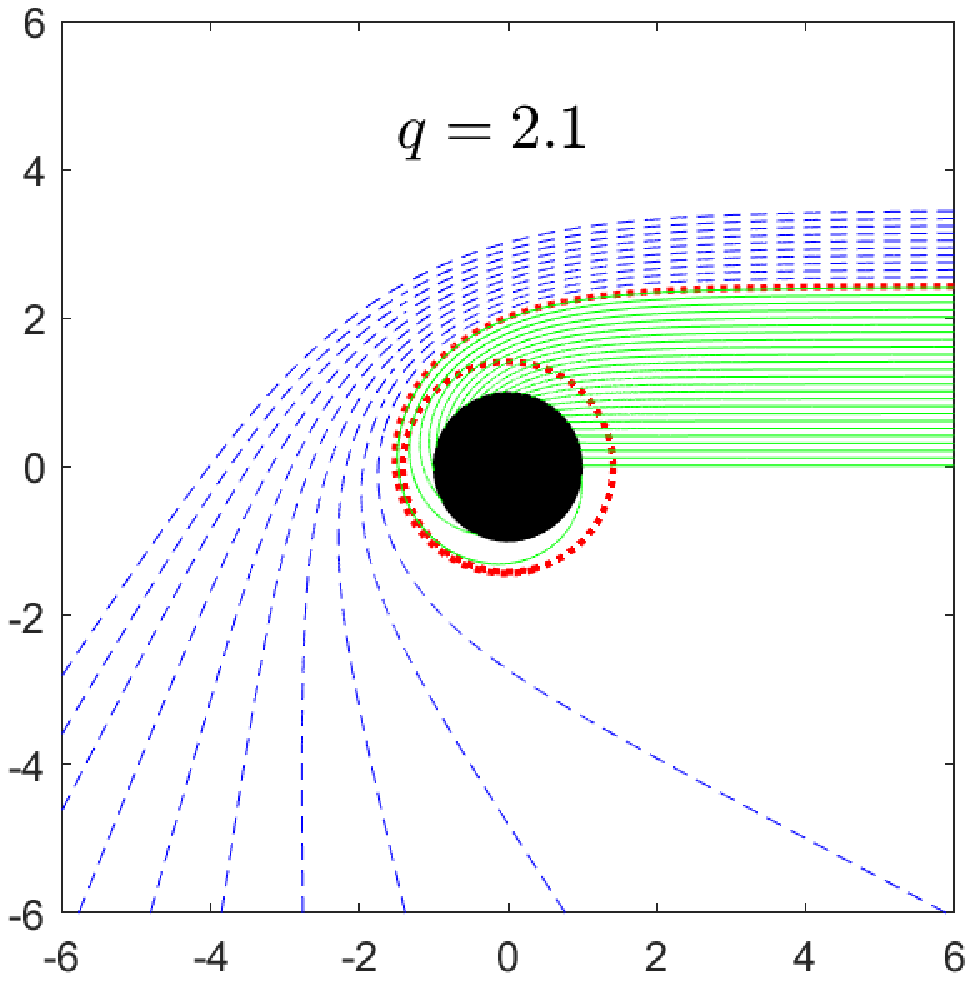}\includegraphics[width=0.45\textwidth]{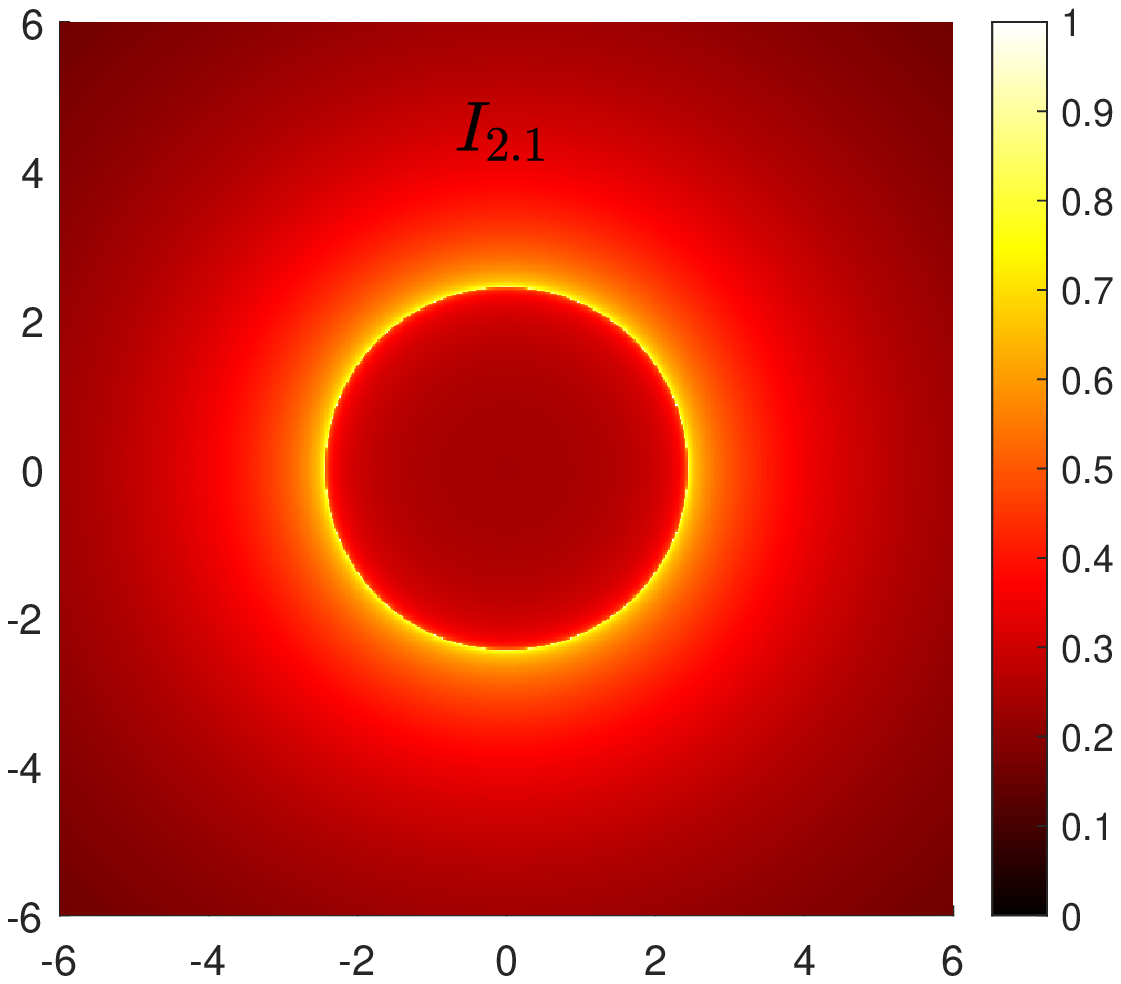}\\
\includegraphics[width=0.45\textwidth]{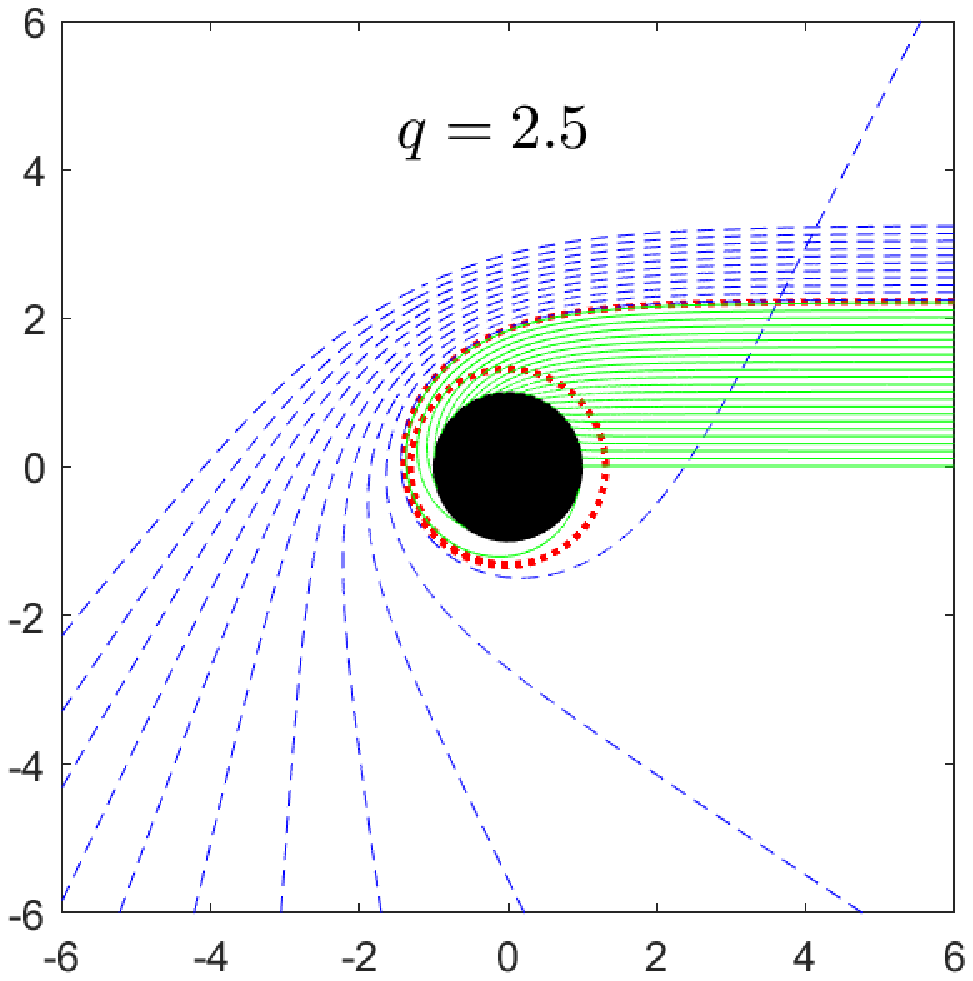}\includegraphics[width=0.45\textwidth]{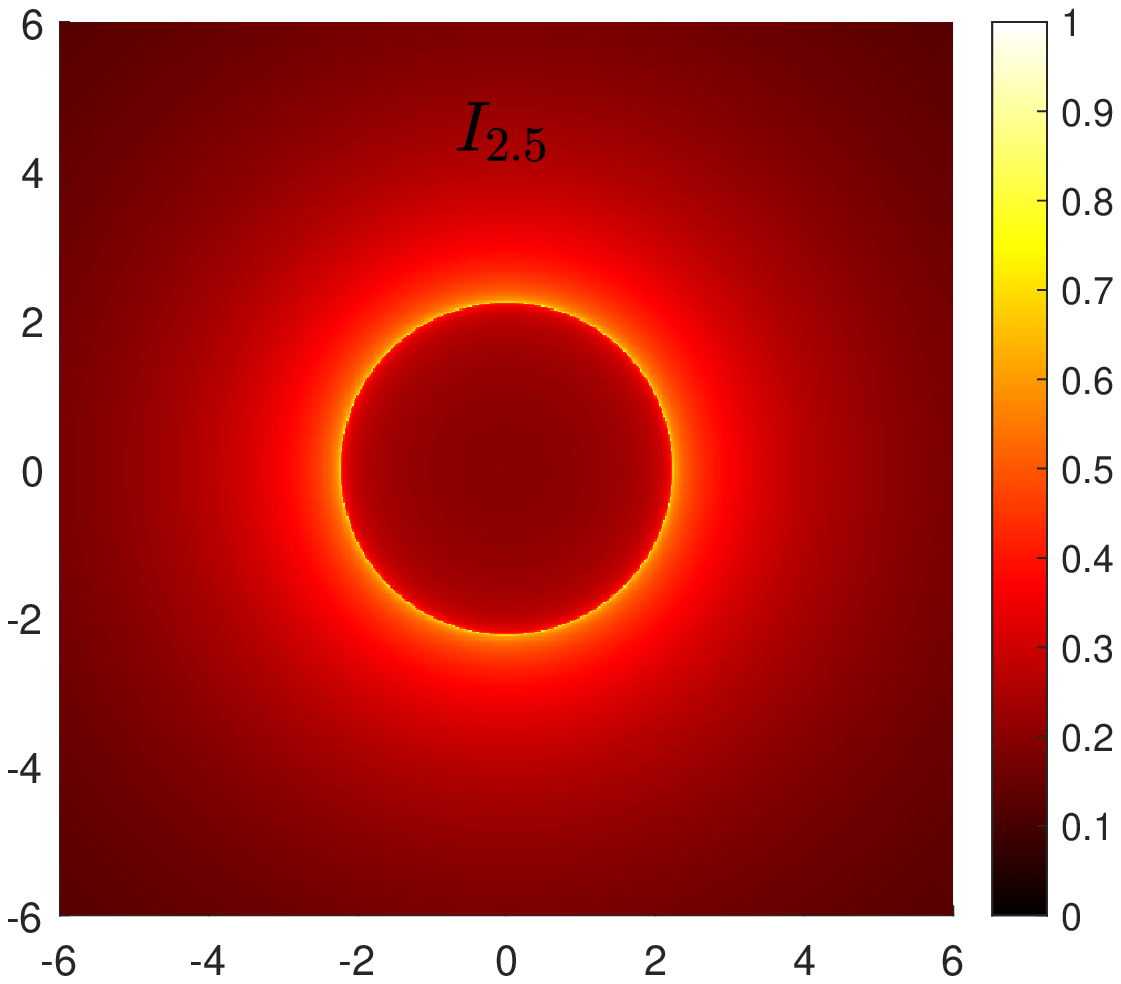}\\
\caption{Light ray trajectories and shadow cast intensity images. All axes in this figure are rescaled by the areal radius of the throat $\rhom$. We set $q=2.0,2.1,2.5$ from top to bottom panels. Left panels: trajectories of the light rays simulated according to Eq. \eqref{orbitu}. Green solid curves, red dotted curves and blue dashed curves depict trajectories with $b<\bph$, $b=\bph$ and $b>\rph$ respectively. The interval of adjacent orbits is $\Delta b/\rhom=0.1$. Red dotted circles correspond to the photon sphere, and the black disks mark the inner patch of spacetime $m/2<r\leq\rmin$ whose influences have been neglected in our simulation. Right panels: the intensity image observed by a distant observer as simulated according to Eq. \eqref{Iu}. As the value of $q$ increases from $2.0$ to $2.5$, the shadow area decreases apparently.}\label{fig-intl}
\end{figure}

Corresponding to Eq. \eqref{ABD}, we can introduce a new variable $u=m/(2r)$ and rewrite the orbit equation \eqref{orbitr} as
\begin{equation}\label{orbitu}
\frac{du}{d\phi}=u\sqrt{\frac{D(u)}{b^2A(u)}-1}
\end{equation}
with
\begin{eqnarray}
\nonumber\frac{D(u)}{A(u)}&=&\frac{m^2}{4}\frac{\left(1+u\right)^{2q+2}}{u^2\left(1-u\right)^{2q-2}}\\
&=&\frac{\rhom^2}{q^2-4}\left(\frac{q-2}{q+2}\right)^{q/2}\frac{\left(1+u\right)^{2q+2}}{u^2\left(1-u\right)^{2q-2}}.
\end{eqnarray}
After fixing the value of minimal areal radius $\rhom$, we solve the trajectories numerically for $q=2.0,2.1,2.5$ and plot them in the left panels of Fig. \ref{fig-intl}. The light rays can be classified into three types \cite{Luminet:1979nyg}: (i) A critical light ray has a critical value of impact parameter $b=\bph$ as given analytically by Eq. \eqref{bc}. It propagates in an unstable circular orbit of radius $\rhoph$ and has a chance to escape to infinity under perturbations. In every left panel of Fig. \ref{fig-intl}, we plot such a critical light ray as a red dotted curve. (ii) Light rays with $b<\bph$ can penetrate the photon sphere. As illustrated by green solid curves in Fig. \ref{fig-intl}, all orbits of these rays start near the throat. (iii) Light rays with $b>\bph$ are always outside the photon sphere. As represented by blue dashed curves in Fig. \ref{fig-intl}, each orbit of such rays is symmetric with respect to a diametrical line through its pericenter.

For a distant observer in the spacetime \eqref{OMW}, $x^2+y^2=R^2=b^2$ in the observed image, the intensity Eq. \eqref{I} can be put in the form
\begin{equation}\label{Iu}
I_q(b)=\int_{\mathrm{ray}}\frac{A(u)^{3/2}}{uD(u)^{1/2}}\sqrt{1+u^2\left(\frac{d\phi}{du}\right)^2}du,
\end{equation}
where $d\phi/du$ is the reciprocal of Eq. \eqref{orbitu}, and
\begin{eqnarray}
\nonumber\frac{A(u)^{3/2}}{uD(u)^{1/2}}&=&\frac{2}{m}\frac{\left(1-u\right)^{2q-1}}{\left(1+u\right)^{2q+1}}\\
&=&\frac{\sqrt{q^2-4}}{\rhom}\left(\frac{q+2}{q-2}\right)^{q/4}\frac{\left(1-u\right)^{2q+1}}{\left(1+u\right)^{2q-1}}.
\end{eqnarray}
We perform this integration using the backward ray shooting method \cite{Luminet:1979nyg}. For light rays with $b<\bph$, the integration is performed from $u=0$ to the throat. For light rays with $b>\bph$, we integrate Eq. \eqref{Iu} from $u=0$ to the pericenter and then to $u=0$. The results for $q=2.0,2.1,2.5$ are illustrated in the right panels of Fig. \ref{fig-intl}, in which the shadow cast shrinks as $q$ increases. This tendency is in agreement with the blue dashed curve in Fig. \ref{fig-rad}.

\begin{figure}
\centering
\includegraphics[width=0.45\textwidth]{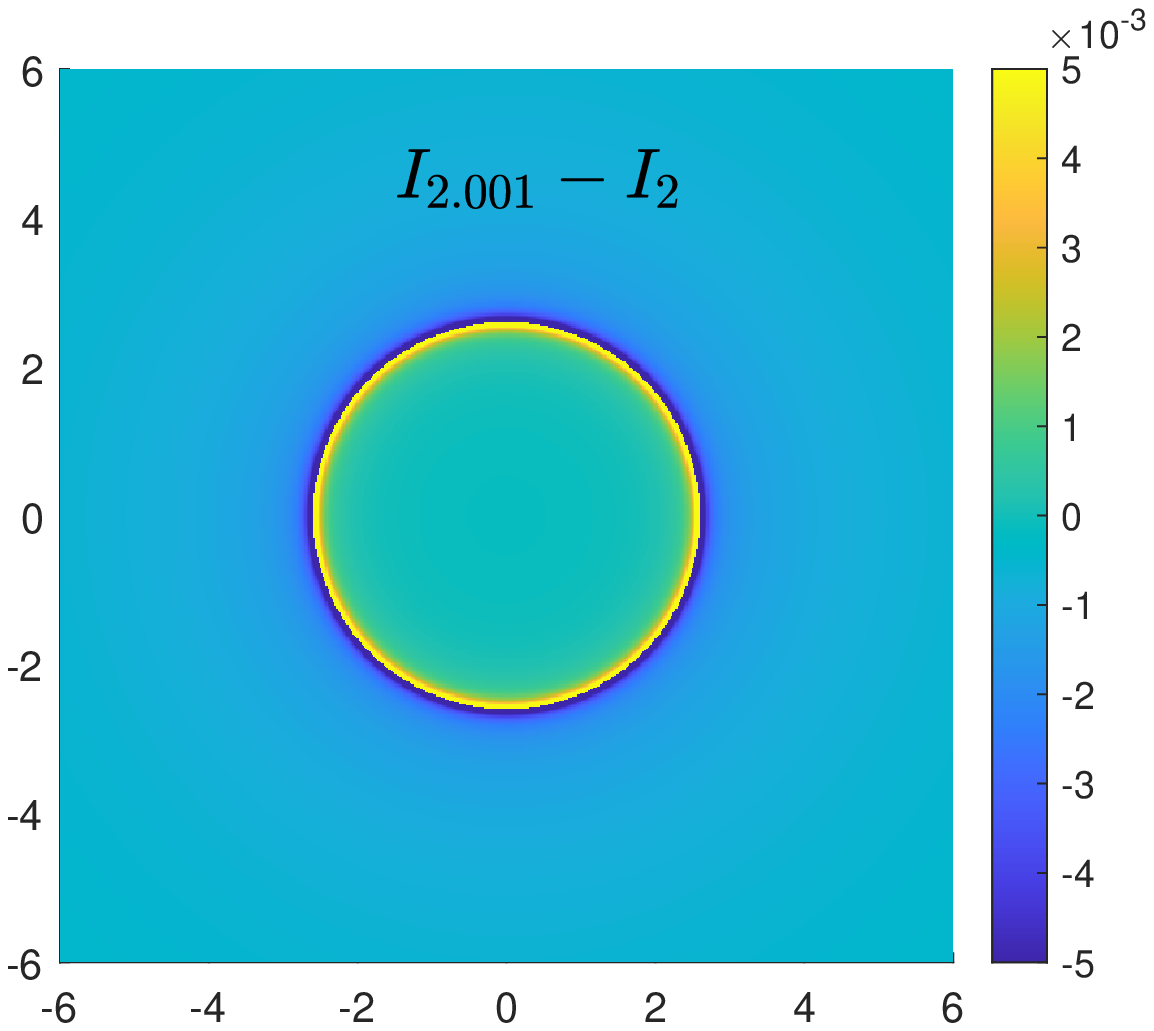}\includegraphics[width=0.45\textwidth]{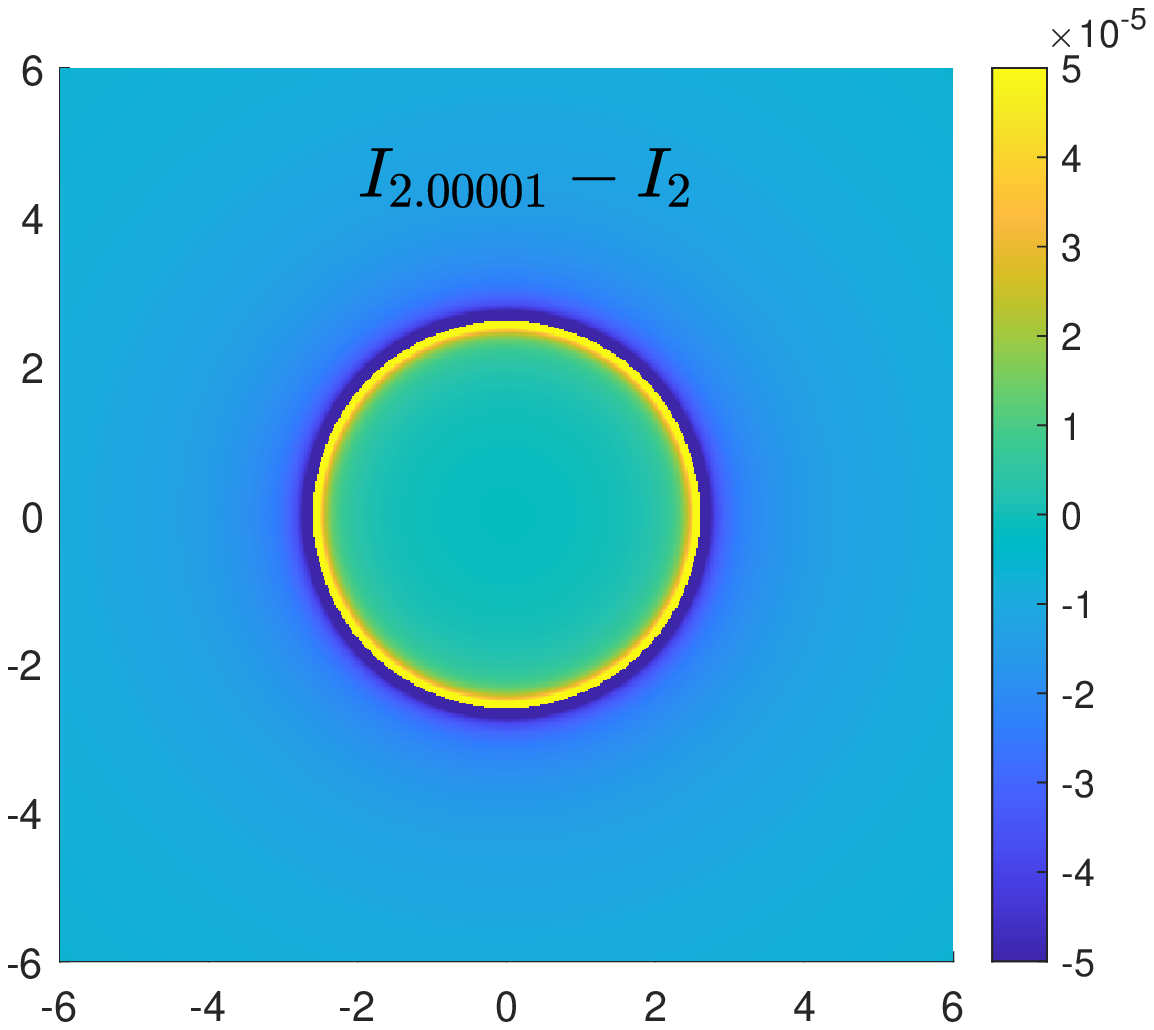}\\
\includegraphics[width=0.45\textwidth]{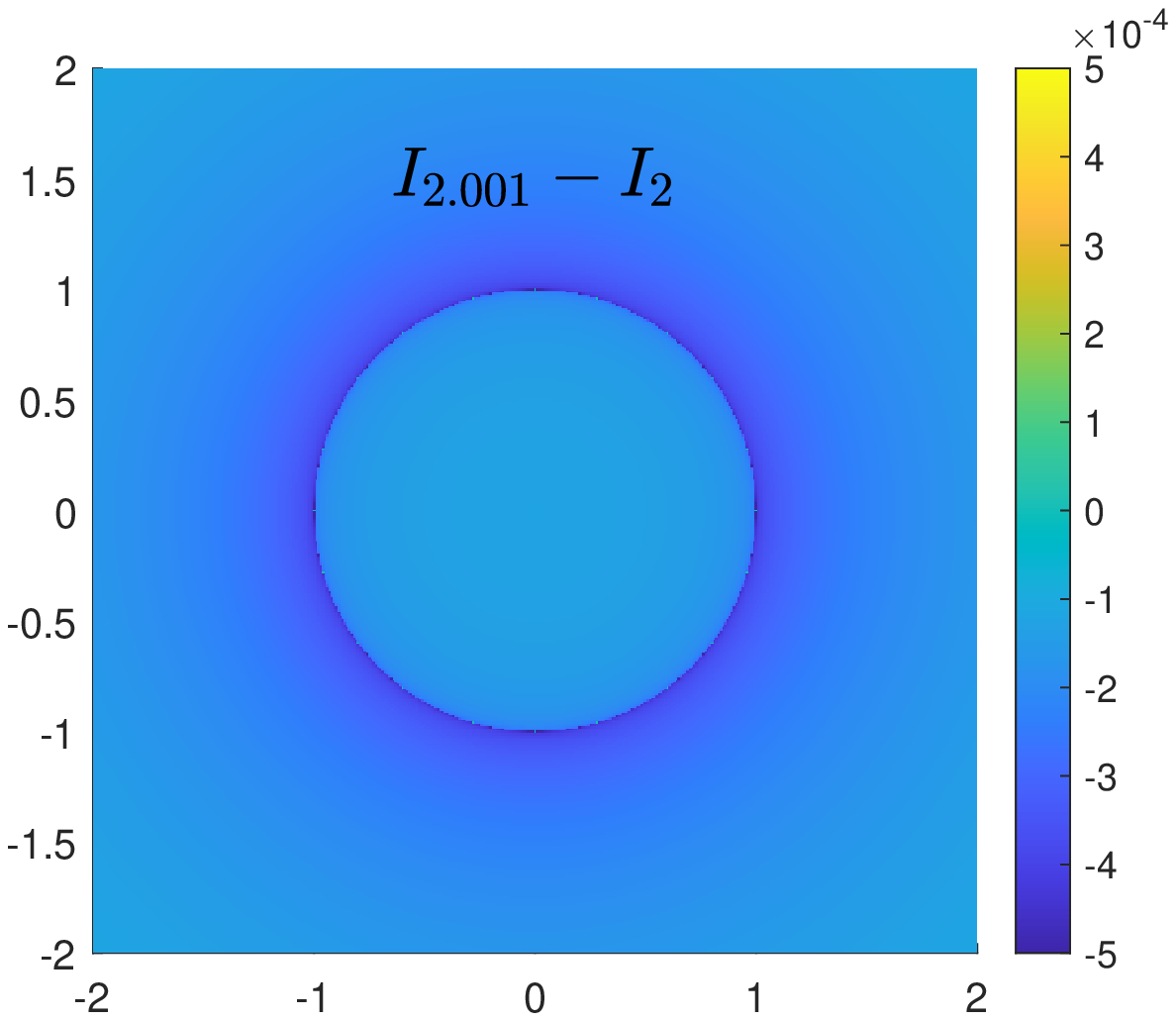}\includegraphics[width=0.45\textwidth]{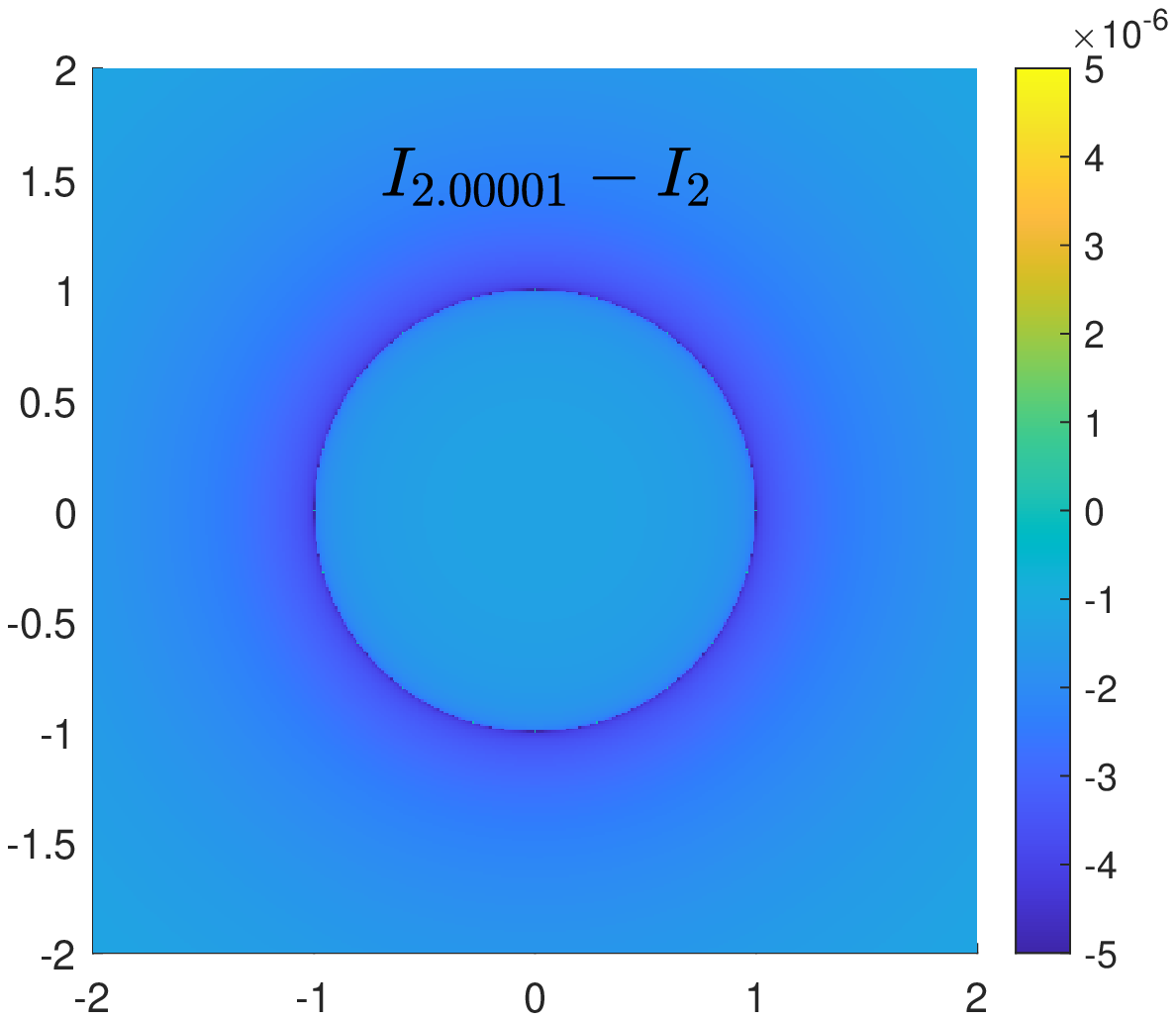}\\
\caption{Intensity difference images for shadows of the wormholelike static aether solutions with the same value of $\rhom$ (upper panels) or $\Rsh$ (lower panels) but different values of $q$. Axes are rescaled by $\rhom$ in upper panels, and by $\Rsh$ in lower panels. The subscript of $I$ is the exact values of $q$ in our simulation.}\label{fig-rints}
\end{figure}

As we have mentioned in Sec. \ref{sect-rev}, the current upper limit on $c_{14}$ is of the order $10^{-5}$. Translated with Eq. \eqref{q}, the upper limit on $q-2$ is of the same order. Therefore, it is interesting to take a closer look at the shadow cast for small values of $q-2$. We have simulated intensity images for $q=2.001,2.00001$, but they are indistinguishable with the image for $q=0$. To make the tiny differences visible, we subtracted the intensity $I_{2}$ from $I_{q}$ and depicted the results in Fig. \ref{fig-rints}. In upper panels, we presented intensity difference images for the shadow casts of solutions with the same value of $\rhom$. They indicate again that the shadow shrinks as $q$ increases. Images in lower panels are for those with the same value of $\Rsh$, indicating that the photon ring becomes less bright as the value of $q$ gets larger.

\subsection{Volume inside the throat}\label{subsect-vol}
In the above, we have neglected light rays traveling out of the throat. The main reason for doing so is that the throat is a marginally trapped hypersurface \cite{Oost:2021tqi}. The other reason, as we want to show in this subsection, is that the volume inside the throat is much smaller than the volume outside as long as $c_{14}$ is small enough.

\begin{figure}
\centering
\includegraphics[width=0.45\textwidth]{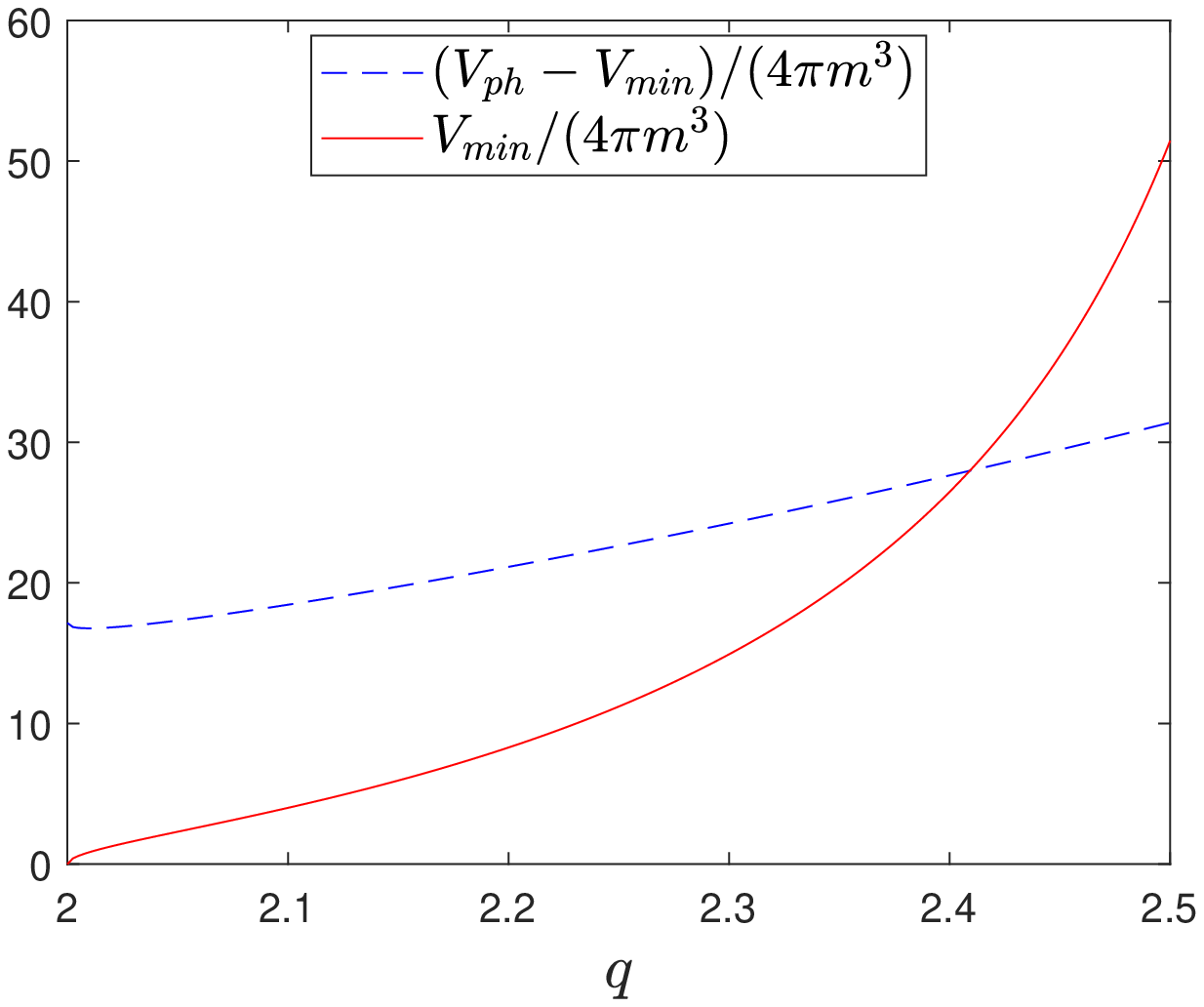}\\
\caption{Simulated volumes according to Eqs. \eqref{vmin} and \eqref{vph}. The red solid line represents the volume inside the throat, while the blue dashed line denotes the volume between the throat and the photon sphere. When $q$ is close to $2.0$, the former is much smaller than the latter.}\label{fig-vol}
\end{figure}

Akin to the volume bounded by the inner horizon of the Reissner-Nordstr\"{o}m black hole,\footnote{In Ref. \cite{Wang:2017zfn}, the second term of Eq. (4) gives the volume bounded by the inner horizon, in which the Killing vector remains timelike.} the volume inside the throat can be calculated with the formula \cite{Christodoulou:2014yia,Wang:2017zfn}
\begin{equation}\label{vol}
V=4\pi\int B(r)^{1/2}D(r)dr,
\end{equation}
which is also applicable to the volume outside the outer horizon of the Reissner-Nordstr\"{o}m black hole or the volume outside the throat here. Inserting Eq. \eqref{ABD} into the integrand and choosing appropriate limits of integral, we get the volume inside the throat
\begin{equation}\label{vmin}
\vmin=4\pi\int_{m/2}^{\rmin}\frac{\left(1+\frac{m}{2r}\right)^{3q/2+3}}{\left(1-\frac{m}{2r}\right)^{3q/2-3}}r^2dr.
\end{equation}
This integral will be divergent if $q\geq8/3$. But for $0\leq q<8/3$, the integration yields a finite result. We perform the integration numerically, and present the results for $2\leq q\leq2.5$ in Fig. \ref{fig-vol} with a red solid curve. Both the areal radius and the proper distance goes to infinity as $r\rightarrow\infty$, therefore the volume outside the throat is infinite. For comparison, we turn to the volume between the throat and the photon sphere
\begin{equation}\label{vph}
\vph-\vmin=4\pi\int_{\rmin}^{\rph}\frac{\left(1+\frac{m}{2r}\right)^{3q/2+3}}{\left(1-\frac{m}{2r}\right)^{3q/2-3}}r^2dr
\end{equation}
which has been worked out numerically and illustrated as a blue dashed curve in Fig. \ref{fig-vol}.

Fig. \ref{fig-vol} shows that the interior volume is significantly smaller than the volume between the throat and the photon sphere when $c_{14}$ is small or equivalently $q$ is close to $2.0$, and they are roughly equal near $q=2.4$. Remind that photons in the shadow casts of black holes are emanated mostly from outside the photon sphere. Therefore, in this section, it is reasonable to omit photons emanated from the inner patch as long as the radiant power density varies smoothly in space.

\section{Conclusion}\label{sect-con}
In Einstein-aether theory, the wormholelike static aether solution is an interesting example of static solution with the aether field parallel to the timelike Killing vector. It was discovered previously in an implicit form \cite{Eling:2006df}. Its explicit analytical form was given recently by Ref. \cite{Oost:2021tqi}. In the present paper, we verified the equivalence of the two forms concretely by a transformation of coordinates and a replacement of implicit functions.

The solution describes a spacetime with a throat like the Schwarzschild wormhole. Its deviation from the Schwarzschild solution is controlled by a parameter $q\geq2$. We studied the motion of photons outside the throat in this spacetime following the customary recipe of calculating black hole shadows. Fixing the area of throat, the shadow of the wormholelike static aether solution shrinks as the parameter $q$ increases. Alternatively, fixing the area of shadow, the brightness of photon ring decreases as $q$ increases.

In this paper, we have paid our attention mainly to the parameter region with $q$ equal to or slightly larger than $2$. This is favored by current theoretical and observational constraints\cite{Oost:2021tqi,Oost:2018tcv} $0\leq c_{14}\leq2.5\times10^{-5}$, or equivalently $0\leq q-2\leq1.3\times10^{-5}$. In this restricted region of parameter, the differences in shadow casts are unnoticeably small. To extract the tiny differences, we plotted the intensity difference images of shadow casts. We also illustrated the influence of $q$ on the shadow cast by plotting the intensity images directly with exaggerated values $q=2.1,2.5$. Both methods lead to the same conclusion above. Due of the smallness of $c_{13}$, our result cannot be tested with the present observational data of the Event Horizon Telescope \cite{EventHorizonTelescope:2019dse,EventHorizonTelescope:2019pgp,EventHorizonTelescope:2019ggy}. In the future, hopefully images of higher resolution and of other black holes will be achieved, then it will be interesting to test our result with more precise observational data.

\end{document}